\documentclass[12pt]{article}
\usepackage{mathrsfs}
\usepackage{amsfonts}
\usepackage{amsmath}
\usepackage{amssymb}
\usepackage{axodraw}
\usepackage{mathptmx}
\usepackage[dvips]{color}
\usepackage{epsfig}
\usepackage{graphicx}
\newcommand{\calL}{{\cal L}}

\newcommand{\calO}{{\cal O}}

\newcommand{\lam}{\lambda}

\newcommand{\eV}{{\rm eV}}
\newcommand{\GeV}{{\rm GeV}}

\newcommand{\TeV}{{\rm TeV}}

\newcommand{\fns}{\footnotesize}

\begin{document}
\baselineskip=16pt

\pagenumbering{arabic}

\vspace{1.0cm}

\begin{center}
{\Large\sf Cascade seesaw for tiny neutrino mass}
\\[10pt]
\vspace{.5 cm}

{Yi Liao\footnote{liaoy@nankai.edu.cn}}

{\it Center for High Energy Physics, Peking University, Beijing
100871, China
\\
School of Physics, Nankai University, Tianjin 300071, China
}

\vspace{2.0ex}

{\bf Abstract}

\end{center}

The accessibility to physics responsible for tiny neutrino mass
suggests that the mass should better originate from certain higher
dimensional operators. The conventional three types of seesaw
operate at dimension five with the help of either a new fermion or
scalar multiplet. Here we propose a seesaw that generates neutrino
mass through a dimension-$(5+4n)$ operator. The seesaw is
functioned by a fermion of isospin $n+1$ and zero hypercharge and
a sequence of scalar multiplets that share unity hypercharge but
have isospin from $\frac{3}{2}$ to $n+\frac{1}{2}$ at a step of
unity. Only the scalar of the highest isospin can couple to the
relevant fermions while only the scalar of the lowest isospin can
directly develop a naturally small vacuum expectation value (VEV).
The VEV is then transmitted to scalars of higher isospin through a
cascading process. No global symmetry is required to forbid lower
dimensional operators. A neutrino mass of desired order can thus
be induced with a relatively low seesaw scale without demanding
too small couplings.

\begin{flushleft}

JHEP keywords: Beyond Standard Model, Neutrino Physics

\end{flushleft}

\newpage

\section{Introduction}
\label{sec:intro}

The tiny neutrino mass can be accommodated in the effective theory
of standard model (SM) by higher dimensional operators. The first
such operator appears at dimension five and is unique,
$\calO_5=\big(\overline{F_L^C}\tilde H^*\big)\big(F_L\tilde
H^*\big)$ \cite{Weinberg:1979sa}, where $H$ and $F_L$ are the
Higgs and left-handed lepton doublets in SM. When $H$ develops a
vacuum expectation value (VEV), $\langle H\rangle$, the operator
yields a neutrino mass of order $m_\nu\sim\lambda\langle
H\rangle^2/\Lambda$. Here $\lambda$ is a product of couplings and
$\Lambda$ a typical heavy mass scale of the underlying high energy
theory that produces the effective operator at low energy.

It has been known for some time that there are exactly three ways to
realize the above operator via tree level interactions of heavy
particles with the Higgs and lepton particles in SM
\cite{Ma:1998dn}. They correspond to the conventional three types of
seesaw mechanisms or underlying theories
\cite{GellMann:1980vs}-\cite{Foot:1988aq}. It would be highly
desirable to discriminate amongst the three theories by looking for
other detectable effects. But this is hard to manage since a sub-eV
neutrino mass generally implies that the new particles are either
extremely heavy or interact with known particles too feebly.

The wayout to the above phenomenological problem is clear from the
point of view of effective theories. The demand for a heavy scale or
small couplings may be alleviated by pushing the neutrino mass
operators to even higher dimensions. Although with a single SM Higgs
field the operators are again unique at each higher dimension
\cite{Liao:2010ku,Liao:2010ny}, one anticipates more possible
underlying theories that can realize the operators. But with a
lowered scale or enhanced couplings one expects to be able to
distinguish them by invoking effects that would otherwise be
unobservable. Roughly speaking, there are two approaches to do so.
In the first approach, one attributes the tiny neutrino mass to a
purely quantum nature arising from radiative effects of heavy
particles. For this purpose one generally employs new particles in
small irreducible representations of the SM gauge group. However, to
forbid the operators to appear at a lower dimension or loop level,
one has to design certain global symmetries that are exact or softly
broken \cite{Zee:1980ai}-\cite{Aoki:2009vf} or quantum numbers like
color \cite{FileviezPerez:2009ud,Ma:2006km} that the SM leptons and
Higgs do not have. In the second approach, one composes new fields
in higher irreducible representations of the gauge group in such a
manner that one has to go through several steps of interactions
between heavy and SM particles to form a neutrino mass operator at
tree level \cite{Babu:2001ex}-\cite{Picek:2009is}. This effectively
pushes up the latter's dimension. In this approach one does not
appeal to global symmetries, but instead chooses representations
judiciously so that low dimensional operators indeed do not occur at
tree level. When they appear at a loop level, they are more
suppressed than those that are available at tree level if the new
particles are not very heavy. It is also possible to combine the two
approaches to new variants \cite{Kanemura:2010bq}. For attempts to
solve the related flavor hierarchy problem using higher dimensional
operators involving scalar fields, see Refs.
\cite{Babu:1999me}-\cite{Giudice:2008uua}.

In this paper we work in the spirit of the second approach. We
attempt to push the neutrino mass operators to higher dimensions
with a minimal set of new heavy particles, though it may be hard
to specify what is minimal. For instance, with a heavy fermion
multiplet of isospin 3 and a heavy scalar multiplet each of
isospin $\frac{3}{2}$ and $\frac{5}{2}$, the operator first
appears at dimension thirteen. At such a high dimension the
tension mentioned above between the tiny neutrino mass and viable
phenomenology would not be a problem.

The paper is organized as follows. In the next section we motivate
our approach by analyzing the quantum numbers of the new fields.
Then we show in sec \ref{sec:vev} how the scalar fields of
increasing isospin develop naturally smaller and smaller VEV's
through cascading interactions with the SM Higgs. The exact form
of neutrino mass is spelt out in the last section together with a
brief mention of possible phenomenology.

\section{Quantum numbers}
\label{sec:quantum}

Our goal is to build with the help of a given set of heavy particles
the lowest dimensional operator that gives neutrino mass upon
spontaneous symmetry breaking in SM. Such an operator must only
involve the SM fields $F_L$ and $H$ and has the unique form,
$\calO_{5+2m}=\calO_5(H^\dagger H)^m$
\cite{Liao:2010ku,Liao:2010ny}. While it requires fields in a higher
representation to push up the operator's dimension, we should
compose a minimal set of fields that can do the job. Since at tree
level gauge fields cannot enter the relevant operators, we restrict
ourselves to new fermions and scalars.

We start with the familiar cases but present them in a manner that
motivates our general analysis. Suppose there is a new scalar
alone. The fermion factor in a potential mass operator must be
$\overline{F_L^C}F_L$, which has the quantum numbers $I=1,~Y/2=-1$
under the SM gauge group. (We suppress the lepton generation index
here but will recover it in the last section.) It would therefore
Yukawa couple to a heavy scalar $\xi$ with $I=1,~Y/2=1$ that would
in turn interact with the SM Higgs $H$ via a trilinear coupling.
The latter induces a VEV of $\xi$ out of that of $H$. Since the
coexistence of the two couplings necessarily breaks lepton number,
their product can be naturally small. This is the type II seesaw.

In the opposite case with a new fermion $\chi$ alone, it must
couple $F_L$ to $H$ and therefore may have $I=0,~1$ and
$Y/2=0,~-1$. (The case with $Y/2=+1$ can be covered using an
appropriately conjugated field.) The choice $Y=0$ corresponds to
the type I and III seesaw respectively. A $\chi$ with $Y/2=-1$
should be vector-like to avoid chiral anomaly. When it is a
singlet, it has the same quantum numbers as the SM lepton singlet
$f_R$ and thus only offers mixing with $f_R$ but not a neutrino
mass. When $\chi$ is a triplet, it couples to $F_L$ and $H$ in the
form, $\bar F_L\tau^aH\chi^a$. In addition to the mixing between
the singly charged fermions, a linear combination of the neutral
components $\chi^0_L$ and $\nu_L$ pairs with $\chi^0_R$ to become
a Dirac fermion, while the orthogonal one remains massless.
Namely, it does not change the numbers of massless and massive
modes. From the viewpoint of seesaw, the lepton number has to be
broken to yield a light Majorana neutrino mass, which however is
not the case here.

The conclusion from the above analysis is that with new scalars or
fermions but not both one cannot get anything else but the
conventional seesaws. A question then naturally arises: what do we
do to go beyond those seesaws, or what is the general case when new
fermion and scalar fields are both present? As this turns out to be
too broad a question with many answers, we focus in what follows on
the more specific case when there is one new fermion $\Sigma$ with
quantum numbers $(I_\Sigma,Y_\Sigma)$ and one new scalar $\Phi$ with
$(I_\Phi,Y_\Phi)$. Let us first consider what restrictions should be
imposed on them to arrive at a high dimensional mass operator. For
definiteness, we assume without losing generality $Y_\Sigma\ge 0$
and $Y_\Phi\ge 0$, which can always be arranged with the help of
conjugate fields. The first restriction, called (R1) below, is that
we exclude the choices $(I_\Sigma,Y_\Sigma)=(0,0),(1,0)$ and
$(I_\Phi,Y_\Phi)=(1,2)$ that cover the conventional seesaws, and
$(I_\Sigma,Y_\Sigma)=(0,2)$ which only causes trivial mixing of
charged fermions. Second, $|I_\Sigma-I_\Phi|=1/2$, so that isospin
allows to couple $\Phi$ and $\Sigma$ to $F_L$ (R2). With a single
chain of fermion lines in a seesaw diagram that starts and ends with
$F_L$, the scalar $\Phi$ must develop a VEV if it is relevant to the
mass generation at all. It is a separate issue how $\Phi$ can
develop a VEV, which will be taken up in the next section. This
means that both $I_\Phi$ and $Y_\Phi/2$ are half-integral (integral)
with $0\le Y_\Phi/2\le I_\Phi$, called (R3). Similarly, for $\Sigma$
to be relevant and considering the restriction (R2), both $I_\Sigma$
and $Y_\Sigma/2$ must be accordingly integral (half-integral) with
$0\le Y_\Sigma/2\le I_\Sigma$. There are thus no fractionally
charged particles. The fourth restriction, (R4), comes from the
neutrality in $Y$ that allows (R4a) $Y_\Sigma+Y_\Phi=1$ for the
Yukawa coupling $(F_L\Sigma\Phi)$, or (R4b) $Y_\Sigma-Y_\Phi=1$ for
$(F_L\Sigma\Phi^\dagger)$, or (R4c) $Y_\Sigma-Y_\Phi=-1$ for
$(F_L\Sigma^C\Phi)$. Finally, the lepton number $L$ must be
explicitly broken (R5).

Consider first the case when both $H$ and $\Phi$ can be connected to
$F_L,~\Sigma$ to induce a neutrino mass, as shown in Fig. 1(a). The
above restrictions also apply to the SM Higgs field $H$. (R2) gives
$I_\Sigma=0,~1$, and (R4) requires $Y_\Sigma=0$ for $(F_L\Sigma H)$
or $(F_L\Sigma^C H)$, or $Y_\Sigma=2$ for $(F_L\Sigma H^\dagger)$.
All choices but $(I_\Sigma,Y_\Sigma)=(1,2)$ are excluded by (R1).
Then (R2) implies $I_\Phi=1/2$ or $3/2$. If $I_\Phi=1/2$, we must
have $Y_\Phi=1$, since with $Y_\Phi=0$, neither Yukawa coupling of
(R4a, R4b, R4c) is possible. This however amounts to a second copy
of the SM Higgs with Yukawa coupling $(F_L\Sigma\Phi^\dagger)$, so
that $L$ is conserved. The choice $I_\Phi=1/2$ thus has to be
discarded. For the remaining choice $I_\Phi=3/2$, one may have
$Y_\Phi=1$ for $(F_L\Sigma\Phi^\dagger)$, or $Y_\Phi=3$ for
$(F_L\Sigma^C\Phi)$. Only for the option $Y_\Phi=3$ can one break
$L$ together with the scalar potential. This is exactly the model
suggested in \cite{Babu:2009aq} that corresponds to a dimension
seven seesaw. Our above analysis shows that it is a unique option
for the seesaw shown in Fig. 1(a).

\begin{center}
\begin{picture}(350,100)(0,0)

\SetOffset(20,60)%
\ArrowLine(0,-30)(30,0)\Line(30,0)(90,0)\ArrowLine(120,-30)(90,0)%
\DashLine(0,30)(30,0){3}\DashLine(90,0)(120,30){3}
\Text(-5,-30)[r]{$F_L$}\Text(125,-30)[l]{$F_L$}%
\Text(60,0)[]{x}\Text(45,-7)[]{$\Sigma$}\Text(75,-7)[]{$\Sigma$}%
\Text(-5,30)[r]{$H$}\Text(125,30)[l]{$\Phi$}%
\Text(60,-35)[]{(a)}

\SetOffset(200,60)%
\ArrowLine(0,-30)(30,0)\Line(30,0)(90,0)\ArrowLine(120,-30)(90,0)%
\DashLine(0,30)(30,0){3}\DashLine(90,0)(120,30){3}
\Text(-5,-30)[r]{$F_L$}\Text(125,-30)[l]{$F_L$}%
\Text(60,0)[]{x}\Text(45,-7)[]{$\Sigma$}\Text(75,-7)[]{$\Sigma$}%
\Text(-5,30)[r]{$\Phi$}\Text(125,30)[l]{$\Phi$}%
\Text(60,-35)[]{(b)}

\SetOffset(20,10)%
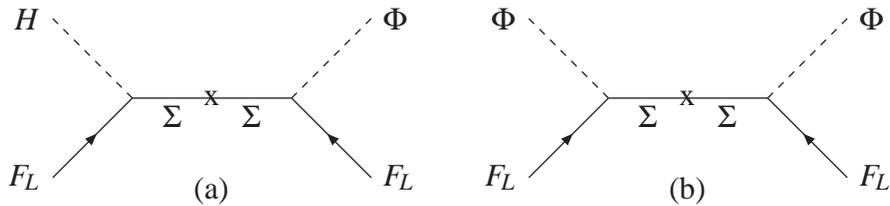
\Text(150,-10)[]{Fig. 1 Seesaw via Yukawa couplings}

\end{picture}
\end{center}

Now we examine the next simplest or more symmetric case in Fig. 1(b)
that will give the new mechanism discussed in this work. Here only
the new scalar $\Phi$ can couple to $(\Sigma,F_L)$ while the SM
Higgs $H$ cannot. First of all, $(I_\Sigma,Y_\Sigma)=(1,2)$ is
excluded to avoid coupling $H$ to $(F_L,\Sigma)$ as in Fig. 1(a),
together with those cases covered in (R1). Next, we choose from (R4)
two forms of Yukawa couplings involving $(\Phi,\Sigma,F_L)$ that
together with the scalar potential will violate $L$. The combination
(R4a,R4b) preserves lepton number and is thus dropped, while the
combination (R4b,R4c) is simply not possible. This leaves us with
the single choice (R4a,R4c) that yields $Y_\Sigma=0$ and $Y_\Phi=1$.
Then (R3) implies that $I_\Sigma$ is integral and $I_\Phi$
half-integral. To avoid the type I and III seesaws, we require
$I_\Sigma\ge 2$ and then $I_\Phi\ge\frac{3}{2}$ from (R2). The
minimal choice is $(I_\Sigma,Y_\Sigma)=(2,0)$ and
$(I_\Phi,Y_\Phi)=(3/2,1)$, which will give a dimension nine mass
operator. This is simpler than the model suggested in
\cite{Picek:2009is}, which has to employ a pair of new scalars with
isospin $3/2$ due to their different choices of hypercharges for the
new fields.

The above analysis for Fig. 1(b) generalizes to arbitrarily high
isospin. The single fermion field $\Sigma$ has quantum numbers
$(I,Y)=(n+1,0)$ with $n\ge 1$, and the scalar of the highest
isospin, $\Phi^{(n+\frac{1}{2})}$, has $(I,Y)=(n+1/2,1)$. To
induce a neutrino mass, $\Phi^{(n+\frac{1}{2})}$ must develop a
naturally small VEV. Since a scalar of isospin $\frac{3}{2}$ can
get a VEV from the $L$-violating term
$\sim\kappa\Phi^{(\frac{3}{2})}\tilde HH\tilde H$ while
$\Phi^{(n+\frac{1}{2})}$ with $n>1$ cannot, the shortest path is
to introduce a sequence of scalar multiplets
$\Phi^{(m+\frac{1}{2})}$ that share the same hypercharge but have
an isospin decreasing at a step of unity, i.e., $1\le m\le n$.
These scalars with intermediate isospin cannot couple to
$(F_L,\Sigma)$ due to too small isospin, but will assist
$\Phi^{(n+\frac{1}{2})}$ to develop a VEV via a cascading process
to be described in the next section.

A few remarks are in order. It would be tempting to ask in this
context why we do not employ two different new scalars in Fig. 1(b).
If their isospins are equal but hypercharges different as in the
model of \cite{Picek:2009is}, this is a matter of simplicity: why
should we introduce one more field when one is sufficient to do the
job? If their isospins are different, the one with a lower isospin
when assigned a correct hypercharge will get a VEV that is less
suppressed as we will show in the next section. Thus the seesaw
employing one scalar with a lower isospin operates at a lower
dimension than the case using two scalars of different isospin, and
thus dominates. Second, a fermion with isospin $n+1$ may couple to
$F_L$ through a scalar of either isospin $n+1/2$ or $n+3/2$. But for
the same reason as described above, the seesaw mass induced from
$\Phi^{(n+\frac{1}{2})}$ dominates since its VEV is less suppressed
compared to that of $\Phi^{(n+\frac{3}{2})}$. Finally, the neutrino
mass operator induced from the symmetric seesaw in Fig. 1(b) jumps
in dimension at a step of four with increasing isospin. To increase
dimension at a step of two it would require several fermion
multiplets and in particular scalar fields of integral isospin that
would develop a VEV through trilinear couplings with the SM Higgs
field. We will not pursue this possibility in the remainder of this
work.

\section{Vacuum expectation values}
\label{sec:vev}

The issue now becomes how a scalar $\Phi^{(n+\frac{1}{2})}$ with
quantum numbers $I=n+\frac{1}{2},~Y=1$ develops a naturally small
VEV. To make our discussion transparent, we start with the
lowest-isospin case, $n=1$, whose scalar potential contains the
terms:
\begin{eqnarray}
V^{(\frac{3}{2})}&\supset&-\mu^2_HH^\dagger
H+\mu^2_\Phi\Phi^\dagger\Phi +\lam_H(H^\dagger H)^2
-\left[\kappa\big(\Phi \tilde HH\tilde H\big)_0
+\textrm{h.c.}\right],
\end{eqnarray}
where $\Phi\equiv\Phi^{(\frac{3}{2})}$ for brevity and the
subscript $0$ denotes the isospin-zero combination of the inside
product. The $\kappa$ term breaks lepton number (together with
Yukawa couplings) and can thus be considered naturally small. It
is important that other parameters are such that $\Phi$ would not
develop a VEV without the $\kappa$ term, in particular
$\mu^2_\Phi>0$. Otherwise it would result in an unwanted massless
Goldstone boson when $\kappa=0$, or a too light scalar when
$\kappa$ is small. The $\kappa$ term and a small
$\langle\Phi\rangle$ have a negligible effect on $\langle
H\rangle$, and cause small mixing between the $\Phi$ and $H$. For
$\mu^2_\Phi\gg\mu_H^2$ and perturbative couplings, we have to good
precision, $\langle H\rangle\approx\sqrt{\mu_H^2/(2\lam_H)}$ which
is assumed real positive without losing generality, and
$\langle\Phi\rangle\approx\kappa^*\langle H\rangle^3/\mu_\Phi^2$.
Note that once a small $\langle\Phi\rangle$ develops from the
$\kappa$ term all other terms only make a subleading correction to
it. These include both $L$-conserving terms like $(\Phi\tilde\Phi
H\tilde H)_0$, $(\Phi\tilde\Phi\Phi\tilde\Phi)_0$, and
$L$-breaking terms like $(\Phi\tilde\Phi\Phi\tilde H)_0$ and
$(\Phi\Phi\tilde H\tilde H)_0$. In other words, the dominant
contribution to $\langle\Phi\rangle$ comes from the $L$-breaking
quartic term that is linear in $\Phi$ and contains as many factors
of $H$ as possible.

When yet another scalar of a higher isospin,
$\Phi^{(\frac{5}{2})}$, is introduced, there will be more quartic
terms in the potential. But most of them are not of our concern
here since they only provide quartic interactions amongst scalars,
and their mass splitting, mixing and trilinear couplings that are
VEV-suppressed. The point here is that we should consider the
largest possible contribution to VEV's for a given set of new
fields. This in turn corresponds to the lowest dimension operator
responsible for neutrino mass that is available in the model.
Since $L$ is necessarily violated by the $\kappa$ term to induce a
small $\langle\Phi^{(\frac{3}{2})}\rangle$, we only need to
consider $L$-conserving ones for all other terms that would induce
a $\langle\Phi^{(\frac{5}{2})}\rangle$. This will give the least
suppressed term in $\langle\Phi^{(\frac{5}{2})}\rangle$. But to
guarantee that it is naturally small, it is again necessary that
the parameters are such that it would vanish if $L$ were not
broken. Since it is not possible to form a $\kappa$-like term for
$\Phi^{(\frac{5}{2})}$, the only way to connect
$\langle\Phi^{(\frac{5}{2})}\rangle$ to $L$ breaking is through an
$L$-conserving quartic term that transfers VEV from
$\Phi^{(\frac{3}{2})}$ to $\Phi^{(\frac{5}{2})}$. As in the case
of $\Phi^{(\frac{3}{2})}$, the term that dominates
$\langle\Phi^{(\frac{5}{2})}\rangle$ should contain as many
factors of $H$ as possible. It is thus linear in both
$\Phi^{(\frac{3}{2})}$ and $\Phi^{(\frac{5}{2})}$, and is unique
upon specifying an equal lepton number,
$-\lam_{(2)}\big(\Phi^{(\frac{5}{2})}\tilde\Phi^{(\frac{3}{2})}
H\tilde H\big)_0+\textrm{h.c.}$. All other terms involve more
factors of the fields $\Phi^{(\frac{3}{2})}$ and
$\Phi^{(\frac{5}{2})}$, and make a correction to the VEV's that is
suppressed by the small VEV's themselves.

The above analysis generalizes obviously to a sequence of scalars,
$\Phi^{(k+\frac{1}{2})}$ with $1\le k\le n$, whose isospin differs
by unity. The terms relevant for consideration of VEV's are
\begin{eqnarray}
V^{(n+\frac{1}{2})}&\supset&-\mu^2_HH^\dagger H%
+\sum_{k=1}^n\mu^2_{(k)}\Phi^{(k+\frac{1}{2})\dagger}
\Phi^{(k+\frac{1}{2})}\nonumber\\
&&+\lam_H(H^\dagger H)^2%
-\sum_{k=1}^n\Big[\lam_{(k)}\big(\Phi^{(k+\frac{1}{2})}
\tilde\Phi^{(k-\frac{1}{2})}H\tilde H\big)_0+\textrm{h.c.}\Big],
\end{eqnarray}
where identifications $\Phi^{(\frac{1}{2})}=H$ and
$\lam_{(1)}=\kappa$ are understood. For a field $\phi$ of isospin
$j$, its conjugate field $\tilde\phi$ that transforms under isospin
exactly as $\phi$ is formed as $\tilde\phi=\tau\phi^*$, where $\tau$
is a matrix with entry $\tau_{m,n}=(-1)^{j-m}\delta_{m,-n}$ ($-j\le
m,~n\le j$) in the eigenstate basis of the third isospin component.
It is evident that the isospin invariant of the quartic term shown
is unique, and can be worked out in terms of the Clebsch-Gordan
coefficients:
\begin{eqnarray}
\big(\Phi^{(k+\frac{1}{2})} \tilde\Phi^{(k-\frac{1}{2})}H\tilde H\big)_0%
=\frac{1}{2\sqrt{2k+1}}\Phi^{(k+\frac{1}{2})}_0
\Phi^{(k-\frac{1}{2})*}_0|H_0|^2+\cdots,
\end{eqnarray}
where the subscript 0 to a field denotes its neutral component, and
the dots stand for the terms not relevant to VEV's. The vanishing
first derivatives of $V^{(n+\frac{1}{2})}$ at VEV's give
\begin{eqnarray}
0&=&\mu^2_{(n)}\langle\Phi_0^{(n+\frac{1}{2})}\rangle %
-\frac{|\langle H_0\rangle|^2}{2\sqrt{2n+1}}
\lam_{(n)}^*\langle\Phi^{(n-\frac{1}{2})}_0\rangle,\\
0&=&\mu^2_{(k)}\langle\Phi_0^{(k+\frac{1}{2})}\rangle %
-\frac{|\langle H_0\rangle|^2}{2\sqrt{2k+1}}\lam_{(k)}^*
\langle\Phi^{(k-\frac{1}{2})}_0\rangle %
-\frac{|\langle H_0\rangle|^2}{2\sqrt{2k+3}}
\lam_{(k+1)}\langle\Phi^{(k+\frac{3}{2})}_0\rangle,
\label{eq_k}
\end{eqnarray}
for $n-1\ge k\ge 1$. The first equation yields
\begin{eqnarray}
\langle\Phi_0^{(n+\frac{1}{2})}\rangle=\frac{1}{2\sqrt{2n+1}} %
\frac{|\langle H_0\rangle|^2}{\mu^2_{(n)}}
\lam_{(n)}^*\langle\Phi^{(n-\frac{1}{2})}_0\rangle,
\end{eqnarray}
which means that $|\langle\Phi_0^{(n+\frac{1}{2})}\rangle|\ll
|\langle\Phi^{(n-\frac{1}{2})}_0\rangle|$ for
$\mu^2_{(n)}\gg|\langle H_0\rangle|^2$ and perturbative couplings.
This implies that the last term in eq (\ref{eq_k}) at $k=n-1$ is
doubly suppressed compared to the first one for
$\mu^2_{(n-1)}\gg|\langle H_0\rangle|^2$, and can be ignored. The
analysis applies to all $k$, so that
\begin{eqnarray}
\langle\Phi_0^{(k+\frac{1}{2})}\rangle=\frac{1}{2\sqrt{2k+1}} %
\frac{|\langle H_0\rangle|^2}{\mu^2_{(k)}}\lam_{(k)}^*
\langle\Phi^{(k-\frac{1}{2})}_0\rangle,~~~n\ge k\ge 1,
\end{eqnarray}
and thus,
\begin{eqnarray}
\langle\Phi_0^{(n+\frac{1}{2})}\rangle=\langle H_0\rangle|\langle
H_0\rangle|^{2n}\prod_{k=1}^n\frac{1}{2\sqrt{2k+1}} %
\frac{\lam_{(k)}^*}{\mu^2_{(k)}}.
\end{eqnarray}
This mechanism of inducing a smaller VEV for a field of a higher
isospin from that of a lower isospin is depicted in Fig. 2 as a
cascading process. Note that the first cascade is suppressed by an
$L$-violating coupling while the sequential cascades are suppressed
by the heavy scalar masses.

\begin{center}
\begin{picture}(360,80)(0,0)

\SetOffset(5,20)%
\SetColor{OliveGreen}%
\DashLine(350,0)(310,0){3}%
\SetColor{Blue}%
\DashArrowLine(310,0)(230,0){3}\DashArrowLine(230,0)(190,0){3}
\DashLine(160,0)(190,0){1}
\DashArrowLine(160,0)(120,0){3}\DashArrowLine(120,0)(40,0){3}
\DashLine(40,0)(0,0){3}%
\SetColor{OliveGreen}%
\DashLine(40,0)(15,40){3}\DashLine(40,0)(65,40){3}%
\DashLine(120,0)(95,40){3}\DashLine(120,0)(145,40){3}%
\DashLine(230,0)(205,40){3}\DashLine(230,0)(255,40){3}%
\DashLine(310,0)(285,40){3}\DashLine(310,0)(335,40){3}%
\Vertex(40,0){2}\Vertex(120,0){2}\Vertex(230,0){2}%
\SetColor{Blue}\Vertex(310,0){2}\SetColor{OliveGreen}

\Text(-5,0)[r]{\Blue{\fns$\Phi^{(n+\frac{1}{2})}$}}
\Text(80,10)[]{\Blue{\fns$\Phi^{(n-\frac{1}{2})}$}}
\Text(150,10)[]{\Blue{\fns$\Phi^{(n-\frac{3}{2})}$}}
\Text(210,10)[]{\Blue{\fns$\Phi^{(\frac{5}{2})}$}}
\Text(270,10)[]{\Blue{\fns$\Phi^{(\frac{3}{2})}$}}

\Text(15,50)[]{\OliveGreen{\fns$\Phi^{(\frac{1}{2})}$}}
\Text(65,50)[]{\OliveGreen{\fns$\Phi^{(\frac{1}{2})}$}}
\Text(95,50)[]{\OliveGreen{\fns$\Phi^{(\frac{1}{2})}$}}
\Text(145,50)[]{\OliveGreen{\fns$\Phi^{(\frac{1}{2})}$}}
\Text(205,50)[]{\OliveGreen{\fns$\Phi^{(\frac{1}{2})}$}}
\Text(255,50)[]{\OliveGreen{\fns$\Phi^{(\frac{1}{2})}$}}
\Text(285,50)[]{\OliveGreen{\fns$\Phi^{(\frac{1}{2})}$}}
\Text(335,50)[]{\OliveGreen{\fns$\Phi^{(\frac{1}{2})}$}}
\Text(355,0)[l]{\OliveGreen{\fns$\Phi^{(\frac{1}{2})}$}}%

\Text(40,-10)[]{\OliveGreen{\fns$\lam_{(n)}$}}
\Text(120,-10)[]{\OliveGreen{\fns$\lam_{(n-1)}$}}
\Text(230,-10)[]{\OliveGreen{\fns$\lam_{(2)}$}}
\Text(310,-10)[]{\Blue{\fns$\lam_{(1)}$}}%

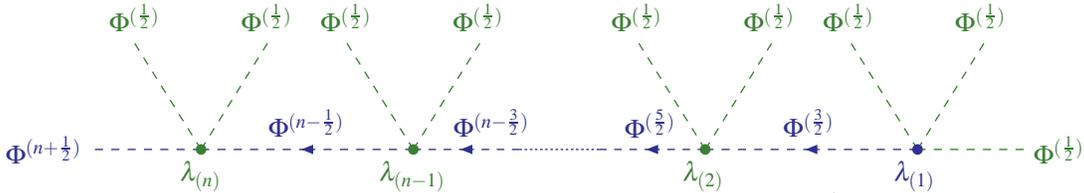
\Text(10,-25)[l]{Fig. 2 VEV's induced via a cascading process. Here
$\Phi^{(\frac{1}{2})}=H$, $\lam_{(1)}=\kappa$.}

\end{picture}
\end{center}

\section{Neutrino mass and discussions}

We are now ready to work out the neutrino mass in a seesaw model
that contains a heavy fermion with $I=n+1,~Y=0$ and a sequence of
heavy scalar multiplets with $Y=1$ and
$I=n+\frac{1}{2},~n-\frac{1}{2},\cdots,\frac{3}{2}$. The bare mass
and Yukawa terms are
\begin{eqnarray}
-\calL_{\textrm{Yuk+mass}}&=&m_\Sigma\overline{\Sigma}\Sigma
+\Big[y_{ij}\overline{F_{Li}}Hf_{Rj}%
+x_j\big(\overline{F_{Lj}^C}\Phi^{(n+\frac{1}{2})}\Sigma\big)_0
+z_j\big(\tilde\Sigma\Phi^{(n+\frac{1}{2})}
F_{Lj}\big)_0+\textrm{h.c.}\Big],
\end{eqnarray}
where $i,~j$ denote the lepton generation. $\tilde\Sigma$ should
be correctly understood as the Dirac-barred conjugate field that
transforms under $SU(2)$ just as $\Sigma$ itself; in the order of
decreasing charge, its components are,
\begin{eqnarray}
\overline{\Sigma}_{-n-1},~-\overline{\Sigma}_{-n},
~\overline{\Sigma}_{-n+1},~\cdots,~-\overline{\Sigma}_{n},
~\overline{\Sigma}_{n+1}.
\end{eqnarray}
The invariant form for a product of three fields $\psi,~\phi,~\chi$
with isospin $n+1$, $n+\frac{1}{2}$, $\frac{1}{2}$ respectively is
unique:
\begin{eqnarray}
(\psi\phi\chi)_0&=&
\frac{1}{\sqrt{(2n+3)(2n+2)}}\sum_{m=-n-1}^{n+1}(-1)^{n+1+m}
\nonumber\\
&\times&
\Big[\sqrt{n+1+m}\psi_{-m}\phi_{m-\frac{1}{2}}\chi_{\frac{1}{2}}
+\sqrt{n+1-m}\psi_{-m}\phi_{m+\frac{1}{2}}\chi_{-\frac{1}{2}}\Big],
\end{eqnarray}
where the subscript to a field denotes its third isospin component.
Applying to our Yukawa couplings, the relevant terms involving only
neutral fields are
\begin{eqnarray}
\big(\overline{F_{Lj}^C}\Phi^{(n+\frac{1}{2})}\Sigma\big)_0&=&
(-1)^{n+1}\frac{1}{\sqrt{2(2n+3)}}
\overline{\nu_{Lj}^C}\Phi^{(n+\frac{1}{2})}_0\Sigma_0
+\cdots, \nonumber\\
\big(\tilde\Sigma\Phi^{(n+\frac{1}{2})}F_{Lj}\big)_0&=&
\frac{1}{\sqrt{2(2n+3)}}\overline{\Sigma_{0}}
\Phi^{(n+\frac{1}{2})}_{0}\nu_{Lj}+\cdots.
\end{eqnarray}

The seesaw neutrino mass can be calculated from Fig. 1(b) with the
above vertices or by solving the equation of motion for the heavy
$\Sigma$ field. We find
\begin{eqnarray}
m^\nu_{jk}&=&(x_jz_k+x_kz_j)\frac{1}{m_\Sigma}(-1)^n\frac{1}{2(2n+3)}
\big(\langle\Phi_0^{(n+\frac{1}{2})}\rangle\big)^2
\nonumber\\
&=&(-1)^n(x_jz_k+x_kz_j)\frac{1}{m_\Sigma}\frac{1}{2(2n+3)} %
\langle H_0\rangle^2\big|\langle H_0\rangle\big|^{4n}
\bigg(\prod_{k=1}^n\frac{1}{2\sqrt{2k+1}} %
\frac{\lam_{(k)}^*}{\mu^2_{(k)}}\bigg)^2,
\end{eqnarray}
which is suppressed by $4n+1$ powers of heavy scales and corresponds
to the neutrino mass operator of dimension $5+4n$, $\calO_{5+4n}$.
Since $m^\nu$ as a matrix in flavor space is a product of two
vectors, there is always one massless neutrino. By judicious gauge
transformations it can be shown \cite{Liao:2006rn,Liao:2010rx} that
there are only two real and one complex physical parameters in the
general complex 3-vectors $x$ and $z$. For instance, in the normal
hierarchy case, one can parameterize without losing generality,
$x^T=(0,0,x)$ and $z^T=(0,z,c_z)$, where $x$ and $z$ are real
positive and $c_z$ is complex. This provides a convenient relation
amongst the neutrino masses, mixing, and the Yukawa couplings. For
an order of magnitude estimate, we ignore the Clebsch-Gordan
coefficients and make the simple-minded approximations: $x\sim z$,
$m_\Sigma\sim\mu_{(k)}\sim M$, and $\lam_{(k)}\sim\lam$ (for $k>1$).
Then the two massive neutrinos have a mass of order, $m\sim
x^2\lam^{2(n-1)}\kappa^2\langle H_0\rangle^{2+4n}M^{-1-4n}$. For
example, with $x\sim 10^{-2}$, $\lam\sim 10^{-1}$, $\kappa\sim
10^{-3}$, and $\langle H_0\rangle\sim 174~\GeV$, the heavy mass is
about $490~\GeV$ at $n=1$ and $190~\GeV$ at $n=2$, to give a desired
neutrino mass around $0.1~\eV$.

The above numerical example also illustrates the point that in
physical applications we do not really need a long chain of
cascading VEV's. With a fermion of isospin $2$ plus a scalar of
isospin $\frac{3}{2}$ (i.e., $n=1$), or at most with a fermion of
isospin $3$ plus a scalar of isospin $\frac{5}{2}$ and a scalar of
isospin $\frac{3}{2}$ ($n=2$), a neutrino mass can be readily
induced at the desired level in the range of parameters that would
be phenomenologically interesting. Introduction of fields of even
higher isospin would over-suppress the neutrino mass. On the other
hand, as we briefly mentioned in the Introduction, a neutrino mass
operator of a lower dimension can be induced at the loop level that
formally amounts to connecting a pair of the $(H,\tilde H)$ fields
in Feynman graphs. A detailed study shows \cite{Ning:2011} that its
contribution is subdominant for a not-too-heavy mass $M$: for
instance, $M<1~\TeV$ at $n=1$, and $M<1.4~\TeV$ at $n=2$, which lies
indeed in the mass range that motivates the current approach.

A salient feature of the seesaw model proposed here is that there
are multiply and equally charged heavy fermions and scalars. They
participate in electroweak gauge interactions and Yukawa couple to
the ordinary leptons. As shown in the above example, these new
particles are not necessarily very heavy, and thus could potentially
be produced at high energy colliders via gauge boson fusion
processes for instance. With enhanced Yukawa couplings with the
ordinary leptons the lightest of them would decay into ordinary
leptons with feasible signatures. On the other hand, the existing
precision data on the rare flavor-changing transitions of the
charged leptons could put stringent constraints on those Yukawa
couplings. Our experience shows (see e.g., \cite{Liao:2009fm} and
\cite{Liao:2010rx}) that it is feasible with the above illustrated
numbers to accommodate the strong constraints in the $\mu e$ sector;
instead, the challenge always rests on whether it is possible to
enhance the rare decays in the $\tau$ sector to a level that would
not be too low compared to the experimental sensitivity available in
the near future.
Furthermore, the new scalars mix with the SM Higgs particle via
interactions in the potential, and thus will modify the production
and decays properties of the latter at high energy colliders. This
could offer additional information on the origin of neutrino mass.
We leave this more comprehensive phenomenological analysis for the
future study \cite{Liao:2011}.

\vspace{0.5cm}
\noindent %
{\bf Acknowledgement}

This work was initiated when I was visiting Institut f\"ur
Theoretische Physik, Universit\"at Heidelberg. I would like to thank
T. Plehn for his kind invitation and support, and the ITP members
for hospitality. I would also like to thank the anonymous referee
and my colleagues for interesting questions and comments that have
helped clarify some issues. The work is supported in part by the
grants NSFC-10775074, NSFC-10975078, NSFC-11025525, and the 973
Program 2010CB833000.

\noindent %

\end{document}